\documentclass[pre,twocolumn,twoside,showpacs,superscriptaddress,floatfix,amsmath]{revtex4}
\usepackage[pdftex]{graphicx,color}

\begin{document}

\title{Conformists and contrarians in a Kuramoto model with identical natural frequencies}

\author{Hyunsuk Hong}
\affiliation{Department of Physics and Research Institute of Physics and 
Chemistry, Chonbuk National University, Jeonju 561-756, Korea}

\author{Steven H. Strogatz}
\affiliation{Department of Mathematics, Cornell University, New York 14853, USA} 
\date{\today}

\begin{abstract}
We consider a variant of the Kuramoto model, in which all the oscillators are now assumed to have the same natural frequency, but some of them are \textit{negatively} coupled to the mean field.  These ``contrarian" oscillators tend to align in antiphase with the mean field, whereas the positively coupled ``conformist" oscillators favor an in-phase relationship.  The interplay between these effects can lead to rich dynamics. In addition to a splitting of the population into two diametrically opposed factions, the system can also display traveling waves, complete incoherence, and a blurred version of the two-faction state.  Exact solutions for these states and their bifurcations are obtained by means of the Watanabe-Strogatz transformation and the Ott-Antonsen ansatz.  Curiously, this system of oscillators with identical frequencies turns out to exhibit more complicated dynamics than its counterpart with heterogeneous natural frequencies.  
\end{abstract}


\maketitle


\section{Introduction}

The Kuramoto model~\cite{Kuramoto} of coupled oscillators has been 
used to shed light on many diverse systems of physical interest, 
particularly those involving synchronization transitions.  
Examples include Josephson junction arrays~\cite{Josephson_junction_arrays}, 
charge-density waves~\cite{charge_density_waves}, laser arrays~\cite{laser_arrays}, 
collective atomic recoil lasers~\cite{atomic_recoil_lasers}, bubbly fluids~\cite{bubbly_fluids},
neutrino flavor oscillations~\cite{neutrino_flavor_oscillations}, electrochemical 
oscillators~\cite{electrochemical_oscillators}, and human crowd behavior~\cite{human_crowd_behavior}.

Originally, however, the Kuramoto model had no known physical applications; these were
only discovered years later.  Kuramoto was led to his model solely by considerations of 
mathematical tractability.  He was seeking an exactly solvable many-oscillator system 
displaying a phase transition to mutual synchronization, in hopes of illuminating this 
novel critical phenomenon seen earlier by Winfree in his simulations of biological 
rhythms~\cite{Winfree}.

In that same spirit, we have begun investigating a family of simple models that 
generalize the Kuramoto model in one key respect: they include \textit{both} positive and negative 
coupling in the same system.  Positive coupling, analogous to a ferromagnetic interaction, tends to align the 
oscillators in phase.  Negative coupling, analogous to an antiferromagnetic interaction, 
drives oscillators apart and favors a phase difference of $\pi$.  When both types of coupling 
are present, the system becomes frustrated.  In this case very little is known about what 
sorts of dynamics and equilibrium states might follow.

Even the mean-field version of such systems remains mysterious.  Twenty years ago, in pioneering 
work, Daido found evidence that Kuramoto models with mixed positive and negative coupling
could undergo a glass transition~\cite{Daido}, but the existence and properties of such 
an ``oscillator glass'' remain unclear~\cite{unclear_properties_of_oscillator_glass}.  Other models with 
mixed attractive/repulsive interactions have since been explored by several authors, who were also motivated by 
analogies to spin glasses, as well as to neural networks with mixed excitatory and inhibitory connections~\cite{differentways_coupling}.  
In each instance it has been difficult to understand the behavior of these models because of their inherent nonlinearity, quenched random interactions, and large numbers of degrees of freedom.  

Inspired by Kuramoto's success in explaining Winfree's synchronization transition 
through the use of a toy model, we wondered whether Daido's oscillator glass transition 
might be similarly rationalized by studying much simpler models with mixed coupling.  
In this paper we analyze the behavior of one such model and find, unfortunately, that this 
particular simplification does not exhibit an oscillator glass.  
Nevertheless, this negative result still provides valuable information.  It shows that 
certain types of frustration are insufficient to produce an oscillator glass, and thereby 
constrains the possible mechanisms at work. 

Furthermore, the model does display some interesting new dynamical phenomena, as we 
discuss below.  And although we are unaware of any physical realization of the model studied 
here, we suspect that such realizations may exist, given the model's simplicity, and given 
the history of the Kuramoto model itself, whose physical relevance was established only 
after the model had been proposed on theoretical grounds.

The governing equations for the model are
	\begin{equation}
	\dot{\phi}^{(s)}_j = \omega + \frac{K_s}{N} \sum_{k=1}^N \sin(\phi_k - \phi^{(s)}_j),
	~~j = 1, \ldots, N
	\label{model}
	\end{equation}
where $\phi^{(s)}_j$ is the phase of the $j$th oscillator in 
the $s$-subpopulation, $\omega$ is its natural frequency, 
and $N$ is the total number of 
oscillators.  The oscillators in subpopulation 1 are assumed to have positive coupling ($K_1 > 0$) to all the other oscillators in the system, whereas those in  
subpopulation 2 have negative coupling $(K_2 < 0)$.  

Equation~(\ref{model}) differs from the classic Kuramoto model in that the distributed natural frequencies $\omega_j$ have been replaced by a uniform natural frequency $\omega$, and the single positive coupling constant $K$ has been replaced by a two-valued coupling constant $K_s$.  In an earlier paper, we considered the case in which $\omega_j$ was kept heterogeneous~\cite{HS_PRL}.  As we will see here however, the long-time dynamics is actually more complicated for the homogeneous case.  This finding is consistent with previous studies of identical oscillators (see Refs.~\cite{Ott_Antonsen,Marvel_Strogatz} for example).

What is unusual about this model is that its pairwise interactions need not be symmetric.  For example, oscillator $k$ could be coupled positively to oscillator $j$ while $j$ is coupled negatively to $k$ in return.  This leads to a novel type of frustration.  Although unfamiliar, it may be physically realizable in certain kinds of series arrays of Josephson junctions~\cite{Josephson_junction_arrays} or in liquid crystal spatial light modulators suitably coupled by global optoelectronic feedback~\cite{rogers2004}.  Because of its asymmetry, this form of coupling is non-variational; no energy function exists, and the dynamics do not correspond to relaxation or gradient descent down an energy landscape.   While this might seem unnatural in some physical settings (e.g., magnetic spin systems), it is more plausible in certain social or political contexts.  In particular, if we set $\omega = 0$ (as can be done without loss of generality by going into a suitable rotating frame, or equivalently, by replacing $\phi_j$ with $\phi_j + \omega t$), the model starts to resemble some of the existing models of social opinion 
formation~\cite{opinion_formation}.  

To see the connection, imagine a spectrum of opinions or attitudes that can be laid out as points on a circle, rather than as points on a line.  For instance, Binmore~\cite{Binmore} has argued that political attitudes are more properly represented this way than as the usual linear continuum from left wing to right wing.  Now consider a population of indifferent individuals who have no preferred phase along the circle, or, to continue the analogy, no inherent political preference.  All that matters to them is what other people think.  Such an individual updates his or her political ``phase" continuously, based on where he or she stands in relation to the prevailing sentiment.  Some individuals -- the conformists -- want to be in phase with conventional wisdom, whatever it happens to be, whereas contrarians reflexively oppose it.  

The question is: depending on the relative proportions of conformists and contrarians, and depending on how intensely they react to the prevailing opinion, what will this population do in the long run?  Split into two camps?  Fail to reach any consensus at all?  Or cycle through all attitudes periodically?  As we'll see, all of these are possible long-term outcomes, depending on the choice of model parameters.

\begin{figure}[t]
	\includegraphics{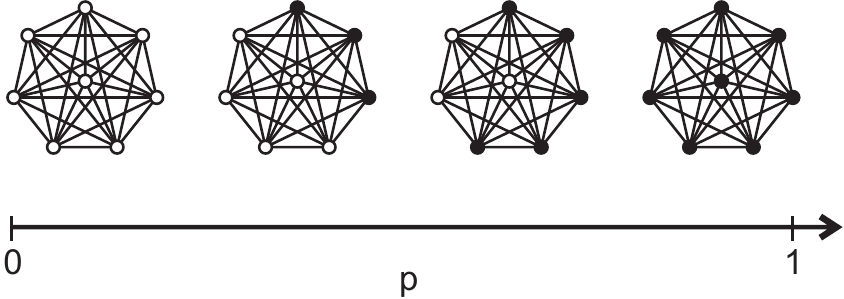}
	\caption{\label{fig:network} A schematic plot of a set of eight identical 
oscillators with positive and negative coupling.  Open circles represent 
oscillators with negative coupling to all the others; filled circles denote 
oscillators with positive coupling.  When $p = 0$, all oscillators are contrarians.  
However, as $p$ is increased, conformists begin to appear and eventually replace 
all contrarians as $p$ approaches $1$.}
	\end{figure}

Let $p$ denote the fraction of oscillators with positive coupling; thus the system consists of $pN$ conformists and 
$qN$ contrarians, where $q = 1-p$ (Fig.~\ref{fig:network}).
When $p=0$, all the oscillators repel one another, a case
explored in Ref.~\cite{repulsive_synch_Tsimring}.

In what follows, we will examine the dynamics of this system as its parameters are varied.  We will continue to use the metaphorical
language of conformists and contrarians, although perhaps we should stress that we do not intend the model to be taken literally as 
a description of real social situations.  It is a toy model.  Like the Kuramoto model itself, it is being offered on theoretical grounds, without any particular physical realization in mind.  The goal is to clarify the dynamical consequences of mixed coupling, by investigating a particularly simple and tractable special case.  Our hope is that such an investigation may bring us a step closer toward solving the puzzle of ``oscillator glass"~\cite{Daido, unclear_properties_of_oscillator_glass, differentways_coupling}.

\section{Dimensional Reduction}

The dynamical system given by Eq.~(\ref{model}) enjoys a highly non-generic structure.  It has $N-6$ constants of motion, for all $N > 6$.  In geometrical terms, its phase space is foliated by an $(N-6)$-parameter family of six-dimensional invariant manifolds. 

These results can be seen explicitly by using a theoretical device discovered by Watanabe and Strogatz~\cite{WS94}, and recently generalized by 
Pikovsky and Rosenblum~\cite{Pikovsky_Rosenblum}.    
Rewrite Eq.~(\ref{model}) as  
	\begin{equation}
	\dot{\phi}^{(s)}_j = f + K_s g\cos \phi^{(s)}_j + K_s h\sin \phi^{(s)}_j,
	\label{fgs}
	\end{equation}
where 
$f=\omega$, $g = (1/N)\sum_{k=1}^{N}\sin\phi_k$ and  
$h = -(1/N)\sum_{k=1}^{N}\cos\phi_k$.  As discussed above, assume that $\omega = 0$ without loss of generality.
Watanabe and Strogatz showed that all solutions $\phi_j(t)$ of Eq.~(\ref{fgs}) can be expressed as 
	\begin{equation}
	\tan\left[\frac{\phi^{(s)}_j(t) - \Theta_{s}(t)}{2}\right] 
	= \sqrt{\frac{1 + \gamma_s(t)}{1 - \gamma_s(t)}} 
	\tan\left[\frac{\psi^{(s)}_j - \Psi_{s}(t)}{2}\right],
	\label{WStf}
	\end{equation}
where the $\psi^{(s)}_j$ in Eq.~(\ref{WStf}) are \textit{constant}, and 
$\gamma_s(t)$, $\Psi_s(t)$, and $\Theta_s(t)$ evolve according to the 
ordinary differential equations
	\begin{eqnarray}
	&\dot{\gamma}_s &= -(1-\gamma_s^2)K_s(g \sin \Theta_s - h \cos \Theta_s), \nonumber \\
	&\dot{\Psi}_s   &= -\frac{\sqrt{1-\gamma_s^2}}{\gamma_s}K_s \Bigg(g \cos \Theta_s + h \sin \Theta_s \Bigg), \label{WSeq} \\
	&\dot{\Theta}_s &= -\frac{K_s}{\gamma_s}\Bigg(g\cos \Theta_s + h\sin \Theta_s \Bigg). \nonumber
	\end{eqnarray}
Here again the variables $\phi^{(s)}_j(t)$ denote the oscillator phases in the 
$s$-subpopulation, where $s = 1, 2$.  The constants $\psi^{(s)}_j$ represent a set of fixed 
phases on which the transformation operates.  For example, if we set 
$\gamma_s(0) = \Psi_s(0) = \Theta_s(0) = 0$, then the $\psi_j$ 
are just the initial phases $\phi_j(0)$.  Since the variables $\gamma_s(t)$, $\Psi_s(t)$, and $\Theta_s(t)$ are the same for all $j$ within each subpopulation, the flow governed by Eq.~(\ref{WSeq}) is effectively 6-dimensional, as claimed.  It describes the dynamics restricted to the invariant manifold labelled by the choice of the constants $\psi^{(s)}_j$, of which $N-6$ turn out to be independent~\cite{WS94}.

A further reduction is possible in the continuum limit $N \rightarrow \infty$, in the special case where the phases $\psi_j$ are uniformly distributed around the circle.   Then the transformation Eq.~(\ref{WStf}) maps old phases $\psi_j$ to new phases $\phi_j$ such that for each subpopulation, a uniform distribution of the $\psi_j$ maps onto a Poisson kernel distribution of 
$\phi_j$~\cite{Marvel_Strogatz,WS94,Pikovsky_Rosenblum,Marvel_Mirollo_Strogatz}.
This implies that the set of states in which each subpopulation is 
distributed like a Poisson kernel is dynamically 
invariant~(see Ref.~\cite{Marvel_Strogatz,Ott_Antonsen,WS94,Pikovsky_Rosenblum,Marvel_Mirollo_Strogatz} for more about this). 
From here on, we will refer to this distinguished invariant manifold as 
the \textit{Poisson submanifold}.  

Incidentally, these considerations underlie the (otherwise seemingly miraculous)
 ansatz discovered by Ott and Antonsen~\cite{Ott_Antonsen}.  They found that 
Poisson kernels are also dynamically invariant for the original 
Kuramoto model, where the oscillator frequencies are non-identical.  
This beautiful invariance property has its origin in 
group theory~\cite{Marvel_Mirollo_Strogatz} and has allowed many new 
insights to be gained into the dynamics of the Kuramoto model and its 
relatives~\cite{Ott_Antonsen, Martens_bimodal, Childs_Strogatz, Carlo_chimera}.

On the Poisson submanifold, two of the equations in the system~(\ref{WSeq}) decouple from the other four.  Thus, as we will see in detail below, the dynamics become effectively four-dimensional there.  And because of an additional rotational symmetry (stemming from the fact that the right hand side of Eq.~(\ref{model})  involves only phase differences, not absolute phases), the flow can be further reduced to a three-dimensional system, which appears later in this paper as 
Eq.~(\ref{eq:OA}).

\section{Simulation of the Reduced System}

We now numerically explore the dynamics of the six-dimensional system given by Eq.~(\ref{WSeq}) to get a sense of its equilibrium states.  To do so, we recall that $p$ is the fraction of the $N$ oscillators that are conformists, and we define $C$ as the relative intensity of the conformist coupling: $C = K_1/(K_1 - K_2)$.  Thus, values of $C$ close to 1 mean the conformists are much more intense in their desire to be like everyone else, as compared to the  relatively mild obstinacy of the contrarians. On the other hand, when $C$ is close to 0, the conformists are tepid while the contrarians are passionate.   

For our initial conditions, we choose each $\gamma_s(0)$, $\Psi_s(0)$ and $\Theta_s(0)$ uniformly at random from $[0,1)$, $[-\pi,\pi)$ and $[-\pi,\pi)$, respectively.  In addition, we set the $N$ constants $\psi_j$ such that each subpopulation is evenly spaced on the interval $[-a\pi, a\pi)$ for $a \leq 1$.  For example if the $pN$ conformists are indexed first and the $qN$ contrarians after, we set $\psi_j$ to be
	\begin{equation} \label{16}
	\psi_j = \left\{
		\begin{array}{cc}
		\dfrac{2a\pi(j-pN/2)}{pN}    & j = 1, \ldots, pN,  \\
		\vspace{-3pt}                & \vspace{-3pt}       \\
		\dfrac{2a\pi(j-pN-qN/2)}{qN} & j = pN+1, \ldots, N.
		\end{array} \right.
	\end{equation}
As we noted in the previous section, choosing $a = 1$ confines the trajectories to a distinguished submanifold of the phase space in which the new phases $\phi_j$ for each subpopulation are distributed like a Poisson kernel.  Choosing $a < 1$ instead gives an initial condition off this special  manifold and therefore allows the system to explore other parts of phase space~\cite{WS94,Marvel_Hong_Strogatz}.

We begin by numerically integrating Eq.~(\ref{WSeq}) from initial conditions on the Poisson submanifold.  After transients have decayed we compute the final phase density $P(\phi)$ and order parameter $Z$,  defined by 
	\begin{equation}
	Z \equiv R e^{i\theta} = \frac{1}{N}\sum_{j=1}^{N} e^{i\phi_j}.
	\label{Z}
	\end{equation}
We also compute the final order parameters of each subpopulation:
	\begin{equation} 
	Z_s \equiv r_s e^{i\theta_s} = \frac{1}{N_s} \sum_{j \in J_s}e^{i\phi_j},
	\label{Zs}
	\end{equation}
where $J_1 = \{1, \ldots, pN\}$ and $J_2 = \{pN+1, \ldots, N\}$.  Here $r_s$ represents the degree of synchronization of the subpopulation $s$, and $\theta_s$ denotes its average phase.  Likewise, $N_s$ is the number of oscillators in this subpopulation.  The integration itself is done using Heun's method with a time step of $0.01$.
 
	\begin{figure}
	\includegraphics{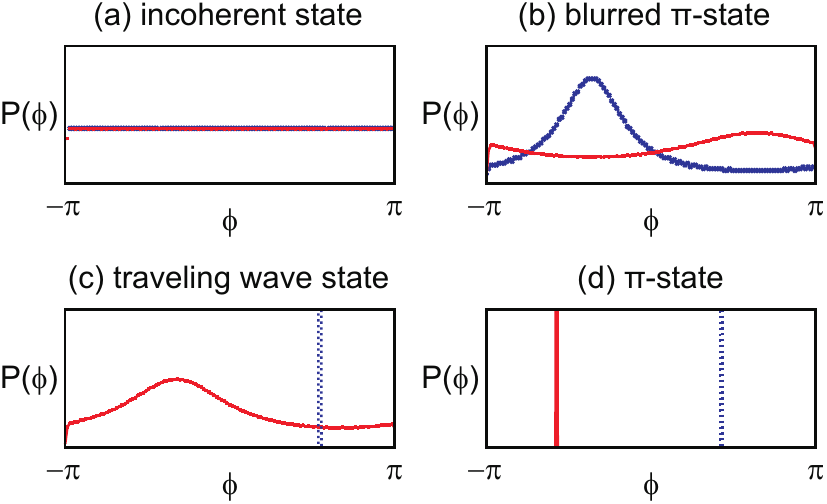}
	\caption{\label{fourstates} (Color online) Four states commonly observed in the long-time behavior of numerical trials on the Poisson submanifold.  The integration was performed for $N = 10^4$ oscillators and the final states are presented as histograms with a bin size of 0.01.  Conformist oscillators are shown in blue and contrarian oscillators in red.  The four states shown are (a) the incoherent state at $(p,C) = (1/20,2/3)$, (b) a blurred $\pi$-state at $(p,C) = (1/4,2/3)$, (c) a traveling wave state at $(p,C) = (1/4,2/3)$ and (d) a $\pi$-state at $(p,C) = (4/5,2/3)$.}
	\end{figure}

Out of the whole $10^5$ time steps, the first $7\times 10^4$ time steps are discarded 
as transients, after which the quantities of interest were measured and averaged for 
the remaining time steps. The system generally seems to end up in one of 
four states (Fig.~\ref{fourstates}):
	\begin{itemize}
	\item[(a)] The \textit{incoherent state}, in which both the conformists and contrarians are uniformly distributed around the unit circle in the complex plane, yielding $r_1 = r_2 = 0$.  In terms of the political analogy discussed earlier, this means that no predominant attitude emerges in the population.  All points on the political spectrum are equally represented.
	\item[(b)] A one-parameter family of  \textit{blurred $\pi$-states}, corresponding to non-uniformly distributed populations of conformists and contrarians on the unit circle.  The peaks of their phase distributions are blurred and separated from one another by an angle of $\pi$.  Here the political interpretation is that two main factions have emerged, in diametrical opposition to one another.  They could lie anywhere on the political spectrum, but once they emerge,  the contrarians oppose the conformist view.  And because of the blurred nature of both peaks, ``fringe'' views are also present on either side of the two main attitudes.  
	\item[(c)] A \textit{traveling wave state}, in which the conformists and contrarians exhibit full and partial phase synchrony, respectively, with the peaks of their phase distributions offset by an angle less than $\pi$.  Here the conformists are unified in their views, yet that consensus view keeps changing, periodically cycling through all possible points on the political spectrum.  Meanwhile the contrarians oppose them, but not quite diametrically, and their opinions remain dispersed throughout. 
	\item[(d)] The \textit{$\pi$-state}, in which the conformists and contrarians are completely synchronized into two antipodal delta functions (and thus $r_1 = r_2 = 1$ and $|\theta_2-\theta_1|=\pi$).  This simple state represents implacable polarization between two unified and unchanging factions.
	\end{itemize}

\begin{figure}
	\includegraphics{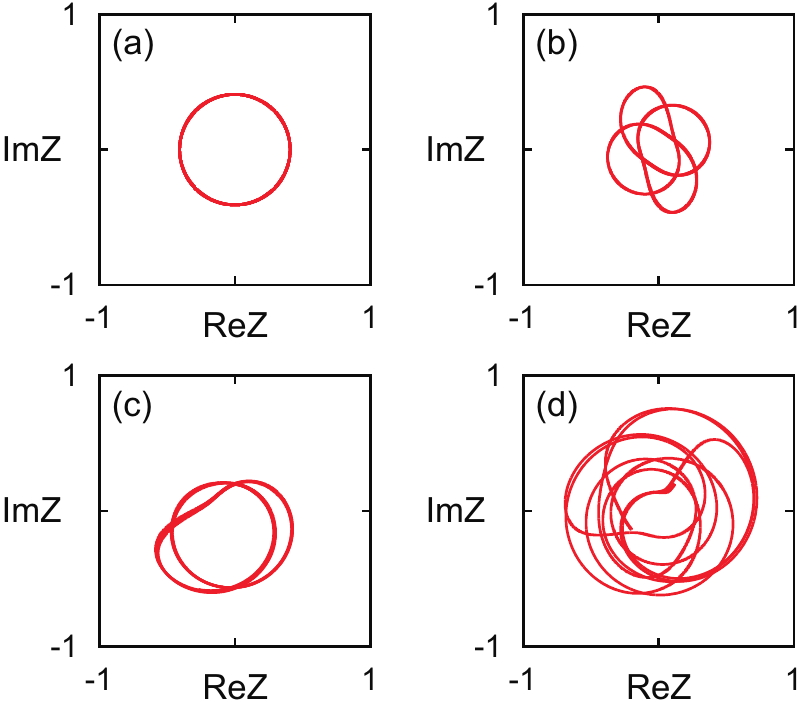}
	\caption{\label{ReZImZ} Trajectories of the order parameter $Z(t)$ both (a) on and (b)-(d) off the Poisson submanifold.  The specific parameter values $(p,C,a)$ for these states are (a) $(1/2,2/3,1)$, (b) $(2/5,2/3,1/2)$, (c) $(1/2,2/3,1/2)$ and (d) $(11/20,2/3,3/10)$.}
	\end{figure}

The offset by an angle less than $\pi$ for the traveling wave state induces a nonzero wave speed.  Hence $Z$ traces out a circular orbit, as shown in Fig.~\ref{ReZImZ}(a).  Interestingly, the traveling wave state has been also found in analogous systems with heterogeneous natural frequencies~\cite{HS_PRL}.

The long-time dynamics of $Z$ becomes substantially more complicated when 
we evenly space the constants $\psi_j$ on the interval $[-a\pi, a\pi)$ for 
$a < 1$, corresponding to initial conditions lying off the Poisson submanifold. 
Typical trajectories appear either quasiperiodic or possibly chaotic in these 
cases, as shown in Figs.~\ref{ReZImZ}(b)-(d).  Similarly non-periodic behavior 
off the Poisson submanifold has been seen in other systems of oscillators 
with identical frequencies~\cite{WS94,Pikovsky_Rosenblum,Marvel_Strogatz}.  

\section{Analysis of the Reduced System}

According to Refs.~\cite{Marvel_Hong_Strogatz,Marvel_Strogatz}, the Watanabe-Strogatz transformation is the real part of the M{\"o}bius transformation.  Additionally, we can convert $\gamma_s$ in Eq.~(\ref{WSeq}) to $r_s$ via the relation $\gamma_s = -2r_s/(1+r_s^2)$ to obtain
	\begin{eqnarray}
	&\dot{r}_s      &= \frac{1-r_s^2}{2}K_s {\rm Re}(Z e^{-i\Theta_s}), \nonumber \\
	&\dot{\Psi}_s   &= \frac{1-r_s^2}{2r_s}K_s {\rm Im} (Z e^{-i\Theta_s}), \label{WSeq_r} \\
	&\dot{\Theta}_s &= \frac{1+r_s^2}{2 r_s}K_s {\rm Im}(Z e^{-i\Theta_s}), \nonumber 
	\end{eqnarray}
where $\text{Re}$ and $\text{Im}$ denote the real and imaginary parts of their 
arguments.  
We here note that the relation between $\gamma_s$ and $r_s$ works only on the Poisson 
submanifold.  It is not satisfied elsewhere~(see details in 
Ref.~\cite{Marvel_Strogatz}). 
Using the fact that $Z = pZ_1 + qZ_2$ and defining 
$\delta = \theta_2 - \theta_1$, Eq.~(\ref{WSeq_r}) becomes
	\begin{eqnarray}
	&\dot{r_1}    &= C(1-r_1^2)(p r_1 + q r_2 \cos\delta), \nonumber \\
	&\dot{r_2}    &= -(1-C)(1-r_2^2)(p r_1 \cos\delta + q r_2),\label{eq:OA}
\\
	&\dot{\delta} &= \sin\delta \Bigg[p(1-C)\Bigg(\frac{r_1}{r_2}+r_1 r_2\Bigg) - qC \Bigg(\frac{r_2}{r_1}+r_1 r_2\Bigg) \Bigg]. \nonumber
	\end{eqnarray}
By a fixed point analysis of Eq.~(\ref{eq:OA}), 
we can show that the four states found above by simulation are in fact the only generic equilibrium states of the reduced system restricted to the Poisson 
submanifold~\cite{Marvel_Hong_Strogatz}.  We summarize several 
interesting points of this analysis in the remainder of this section and compute the order parameter $R$ for the four different states.  We do the latter by making repeated use of the relation $Z = pZ_1 + qZ_2$.  Decomposed more fully, this is $Z = pr_1e^{i\theta_1} + qr_2e^{i\theta_2}$ or $Z = (pr_1 + qr_2e^{i\delta})e^{i\theta_1}$.

\subsection{Incoherent state} 

We start with the easiest case:  The incoherent state has $r_1 = r_2 = 0$, 
so its order parameter $R$ is zero.  By a linear stability 
analysis~\cite{Marvel_Hong_Strogatz}, we find that this state is stable 
when $p < 1 - C$.  This gives us our first bifurcation value of $p$: 
$p_b = 1 - C$.

\subsection{Blurred $\pi$-states} 

The one-parameter family of blurred $\pi$-states is given by the 
following fixed points of Eq.~(\ref{eq:OA}): 
$\delta = \pi$ and $p r_1 = q r_2$ (where $r_1, r_2 \neq 0$).  By our above equations for $Z$, this implies $R = 0$.   Linear stability analysis then shows that the blurred $\pi$-states farthest from the incoherent state begin to lose stability at $p_a = (1 - \sqrt{2C-1})/2$, whereas loss of stability for the entire set of blurred $\pi$-states occurs as $p$ nears $p_b = 1 - C$.  Hence, there are stable blurred $\pi$-states on the same region that the incoherent state is stable.

\subsection{Traveling wave state}

Next we turn to the traveling wave state.  From Fig.~\ref{fourstates}(c), it is clear that the conformists are fully synchronized $(r_1=1)$ 
for this state, so Eq.~(\ref{eq:OA}) reduces to 
	\begin{eqnarray}
	&\dot{r_2} &= -(1-C)(1-r_2^2)(p\cos\delta + q r_2), \nonumber \\
	&\dot{\delta} &= \sin\delta \Bigg[p(1-C)\Bigg(\frac{1}{r_2}+r_2\Bigg)-2qCr_2\Bigg], 
	\end{eqnarray}
where $\sin\delta \neq 0$.  If we then solve for the fixed points of this system, we obtain:
	\begin{equation}
	r_2 = \sqrt{\frac{p(1-C)}{2qC-p(1-C)}}, \quad \delta = \cos^{-1}(-qr_2/p).
	\label{r2delta}
	\end{equation}
This solution can only exist for $0 < r_2^2 < 1$, which implies that the traveling wave state exists only for $p$ greater than $p_a = (1 - \sqrt{2C-1})/2$ and less than $p_c = C$.  Within this region of existence, we can determine the order parameter $R$ for the traveling wave state using the fact that $Z = pr_1 e^{i\theta_1}+q r_2 e^{i\theta_2}$.  Since by definition $Z = Re^{i\theta}$, we know that $R^2 = Z\bar{Z}$, or
	\begin{equation}
	R^2 = p^2 + 2pqr_2\cos\delta + q^2r_2^2.
	\label{13}
	\end{equation}
Substituting Eq.~(\ref{r2delta}) into Eq.~(\ref{13}) gives
	\begin{equation}
	R = \sqrt{p^2 - \frac{p(1-p)^2(1-C)}{2C - p(1+C)}}, 
	\label{eq:Z_TW}
	\end{equation}
which goes to zero at $p = p_a$, as the numerical data suggests.

\subsection{The $\pi$-state} 

Lastly, we consider the $\pi$-state.  As Fig.~\ref{fourstates}(d) suggests, the subpopulations are both synchronized ($r_1 = r_2 = 1$) and antipodal to each other ($\delta =\pi$).  Substituting $r_1 = r_2 = 1$ and $\delta = \pi$ into $Z = pZ_1 + qZ_2$ yields
	\begin{equation} 
	R = 2p - 1
	\label{eq:Z_pi}
	\end{equation} 
which can only be positive for $p\geq 1/2$.  A more systematic stability analysis of the $\pi$-state shows that it is stable for $p > {\rm max}\{C,1/2\}$~\cite{Marvel_Hong_Strogatz}.  For example, when the conformist coupling is twice the magnitude of the contrarian coupling (e.g. $K_1 = 1$ and $K_2 = -1/2$, and so $C = 2/3$), Eq.~(\ref{eq:Z_pi}) implies that the $\pi$-state is stable for $p$ greater than $p_c = 2/3$.  Interestingly, this agrees with the $p_c$ beyond which the traveling wave state does not exist.  Yet $p_c$ is not a bifurcation point since $\delta$ is always $\pi$ for the $\pi$-state but does not approach $\pi$ for the traveling wave state as $p$ approaches $p_c$.  Instead, the $r_2 = 1$ nullcline corresponding to $\dot{r}_2 = 0$ and a parallel nullcline corresponding to $\dot{\delta} = 0$ approach each other as $p$ approaches $p_c$ until they coincide and form a line of fixed points at $p = p_c$~\cite{Marvel_Hong_Strogatz}. 

	\begin{figure}
	\includegraphics{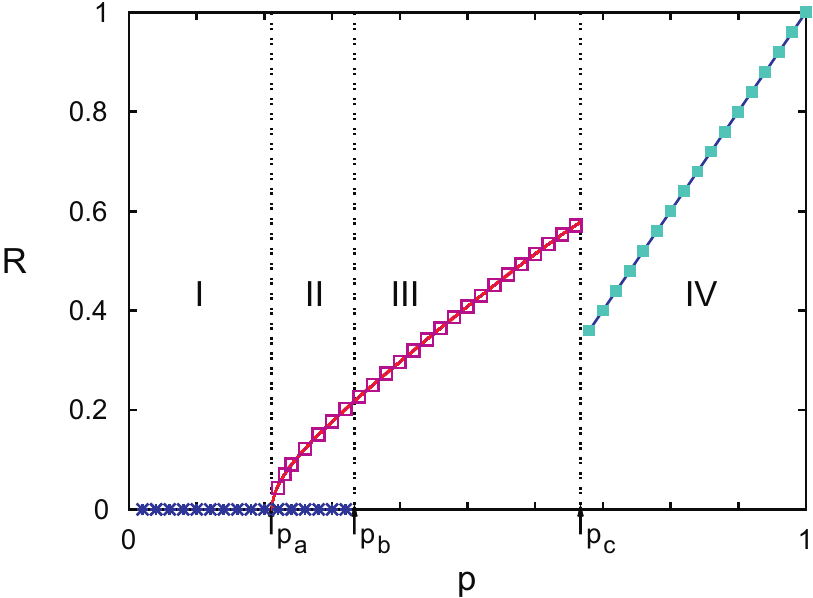}
	\caption{\label{R_p} (Color online) Behavior of the order parameter $R$ for the four different types of equilibrium states as a function of the conformist fraction $p$.  Parameter values: $C = 2/3$, $N = 10^4$ oscillators.  The colored symbols denote $R$ values of fixed points found in numerical simulations of Eq.~(\ref{model}), for initial conditions on the Poisson submanifold.  Blue asterisks denote both the incoherent state and blurred $\pi$-states; open magenta squares denote the traveling wave state; filled cyan squares denote the $\pi$-state.  The solid red curve traces the theoretical value of $R$ for the traveling wave state where it exists, and the blue line gives the theoretical $R$ for the $\pi$-state on its domain of existence.}
	\end{figure}

The above analysis indicates that the stable states of the reduced system reach their boundaries of stability (and sometimes also existence) at three transitional points:  $p_a = (1 - \sqrt{2C-1})/2$, $p_b = 1-C$, and $p_c = {\rm max}\{C,~1/2\}$.  We can verify this by first computing $R$ for the numerically discovered fixed points found at various values of $p$, and then plotting the theoretical curves that we found above on top of these numerical data.  The results in Fig.~\ref{R_p} illustrate the level of agreement between simulation and theory.

We finish by summarizing which states lie in which regions delimited by the transitional points $p_a$, $p_b$ and $p_c$.  In general, there are four regions of behavior, and for $C = 2/3$, their boundaries are $p_a = (1-\sqrt{1/3})/2 \approx 0.21$, $p_b = 1/3$ and $p_c = 2/3$.  On these regions, we have found the following states:
	\begin{itemize}
	\item[I:]{On $0 < p < p_a$, both the incoherent state and all of the blurred $\pi$-states are stable. } 
	\item[II:]{On $p_a < p < p_b$, the incoherent state, some of the blurred $\pi$-states, and the traveling wave state are stable.}
	\item[III:]{On $p_b < p < p_c$, only the traveling wave state is stable.} 
	\item[IV:]{On $p_c < p < 1$, only the $\pi$-state is stable.} 
	\end{itemize}

To be more precise, the notion of stability being used above is that of stability within the Poisson submanifold, not stability within the full phase space.  All states within the Poisson submanifold are neutrally stable to perturbations off the submanifold, because such perturbations carry the system onto another invariant manifold of the foliation discussed earlier.  

\section{Summary}

In this paper, we considered a system of identical oscillators with positive and negative global coupling and investigated how the interplay between the positive and negative interaction affected the collective dynamics and equilibrium states of the system.  We reduced the dynamics of our system from $N$ dimensions to six by means of the Watanabe-Strogatz transformation, and found that in the infinite-$N$ limit there are four types of equilibrium states of the system on a special submanifold of the phase space (the invariant manifold of phase distributions given by Poisson kernels).  Using both numerical and analytical techniques, we characterized each of these equilibrium states, paying particular attention to the illustrative case in which the conformists were coupled twice as strongly to the mean field as the contrarians were.  Even for this slice of parameters, however, we found that a menagerie of complicated states exist throughout the phase space.



\begin{acknowledgments}
H.H acknowledges the hospitality of Cornell University.  
This research was 
supported by the Chonbuk National University (H. H.) and NSF Grants CCF-0835706 and CCF-0832782 (S. H. S.).  We thank Seth Marvel for several useful discussions and the Korea Institute for Advanced Study for providing computing resources for the project.
\end{acknowledgments}


\end{document}